\documentstyle[eqsecnum,preprint,aps,prd,mathrsfs]{revtex}
\def\PB#1#2{\Big\{#1, #2\Big\}}
\def\eijk{\epsilon_{ijk}}
\def\ham{{\mathscr H}}
\def\lag{{\mathscr L}}

\def\half{\frac12}
\def\be{\begin{equation}}
\def\ee{\end{equation}}
\newcommand{\bea}{\begin{eqnarray}}
\newcommand{\eea}{\end{eqnarray}}
\def\we{\approx}

\def\fr{\frac}
\def\a{\alpha}
\def\b{\beta}

\def\d{\delta}
\def\e{\epsilon}

\def\l{\lambda}
\def\m{\mu}
\def\n{\nu}
\def\r{\rho}
\def\s{\sigma}

\def\th{\theta}
\def\w{\omega}

\def\D{\Delta}
\def\d{\delta}

\def\L{\Lambda}

\def\p{\partial}
\def\t{\tilde}
\def\T{\Theta}

\def\no{\nonumber}

\begin{document}
\draft
\preprint{\vbox{\hbox{SBNC/01-02-01}\smallskip
\hbox{hep-th/0102013}}}  

\title{Topologically Massive Non-Abelian Gauge Theories:
Constraints and Deformations} 
\author{E. Harikumar$^1$, Amitabha Lahiri$^2$, M. Sivakumar$^1$ \\ }
\bigskip
\address{$^1$School of Physics, University of
Hyderabad,\\Hyderabad-500 046, INDIA \\
$^2$S.~N.~Bose National Centre for Basic Sciences, \\
Block JD, Sector III, Salt Lake, Calcutta 700 091, INDIA} 

\maketitle

\begin{abstract} 
We study the relationship between three non-Abelian topologically
massive gauge theories, viz. the na\"\i ve non-Abelian
generalization of the Abelian model, Freedman-Townsend model and
the dynamical 2-form theory, in the canonical framework.
Hamiltonian formulation of the na\"\i ve non-Abelian theory is
presented first. The other two non-Abelian models are obtained by
deforming the constraints of this model. We study the role of the
auxiliary vector field in the dynamical 2-form theory in the
canonical framework and show that the dynamical 2-form theory
cannot be considered as the embedded version of na\"\i ve
non-Abelian model. The reducibility aspect and gauge algebra of the
latter models are also discussed.

\end{abstract}
\bigskip

\section{Introduction}\label{intro}
Construction and study of gauge invariant theories of massive
vector fields has been a problem of great intrinsic interest. Such
theories also have a potential application because the Higgs
particle, needed for giving masses to gauge fields and fermions in
the standard model, does not yet have experimental support.
Consequently, alternative theories which have no residual Higgs
scalar, for both Abelian and non-Abelian gauge fields, deserve
closer attention.  One of the oldest models in which there is no
residual scalar particle is the St\"uckelberg formulation.  Another
gauge invariant model where massive gauge fields appear couples a
vector field to a second rank anti-symmetric tensor field through a
topological ${B\wedge F}$ term. The Abelian theory is described by
the Lagrangian
\cite{{Aurilia:1981xg},{Govindarajan:1982jp},{Allen:1991gb}}
\begin{eqnarray}
{\lag}= -\fr{1}{4}F_{\m\n}F^{\m\n} +
\fr{1}{12}H_{\m\n\l}H^{\m\n\l}
+\fr{m}{4}\e_{\m\n\l\s}F^{\m\n}B^{\l\s}, 
\label{intro.alag}
\end{eqnarray}
where $H_{\m\n\l}= \p_\m B_{\n\l} +~{\rm cyclic~terms}$.  This
Lagrangian is invariant under two independent gauge
transformations,
\begin{eqnarray}
\d A_\m &=& \p_\m \w\,, \nonumber \\ 
\d B_{\m\n} &=& (\p_\m \l_\n-\p_\n\l_\m)\,.  
\label{intro.sym}
\end{eqnarray}
The equations of motion following from the
above Lagrangian are
\begin{eqnarray}
\p^\nu F_{\m\n}-\fr{m}{6}\e_{\mu\nu\l\s} H^{\nu\lambda\sigma} &=& 0,
\nonumber \\  
\p^\lambda H_{\m\n\l} - \fr{m}{2}\e_{\mu\nu\lambda\sigma}
F^{\lambda\sigma} &=& 0, 
\end{eqnarray}
which are like the London equations of superconductivity, and
has the interpretation of a massive vector. The massive nature of
the vector boson can also be brought out by summing over
propagators, which leads to the appearance of a pole in the vector
propagator~\cite{{Allen:1991gb}}.

Making a non-Abelian model of massive vector bosons using this
mechanism is a non-trivial task. Na\"\i vely, one can replace the
ordinary derivative $\partial_\mu$ with the gauge-covariant derivative
$D_\m$ in the Lagrangian of Eqn.~(\ref{intro.alag}) to get
\begin{equation}
{\lag} =-\fr{1}{4}F_{\m\n}^{a} F^{a\m\n} + \fr{1}{12} H_{\m\n\l}^{a}
H^{a\m\n\l} +\fr{m}{4}\e_{\m\n\l\s}B^{a\m\n}F^{a\l\s}\,, 
\label{intro.naive}
\end{equation}
where $B$ lives in the adjoint representation of the gauge group
and $H_{\m\n\l}^{a} = (D_\m B_{\n\l})^{a} + {\rm cyclic ~terms~}.$
But although this non-Abelian model is invariant under the usual
gauge transformations
\begin{equation}
\d A_{\m}^a =(D_{\m}\w)^a\,, \qquad \delta B^a_{\mu\nu} = gf^{abc}
B^b_{\mu\nu} \omega^c\,,
\label{intro.gaugesym}
\end{equation}
unlike the Abelian model it is not invariant under the non-Abelian
vector gauge transformations of the 2-form,
\begin{equation}
\d B_{\m\n}^a=(D_\m \l_\n-D_\nu\l_\m)^a\,.
\label{intro.navsym}
\end{equation}
The absence of the vector gauge symmetry makes perturbative
calculations from the Lagrangian of Eqn.~(\ref{intro.naive}) quite
problematic.  Gauge-fixing for $B^a_{\mu\nu}$ is not needed in the
absence of the symmetry, but the quadratic part of the kinetic term
for $B^a_{\mu\nu}$ cannot be inverted without a gauge-fixing term.

The problem runs even deeper, and has been the topic of a recent
`no-go' theorem \cite{{Henneaux:1997mf}} which uses consistent
deformation of the master equation in the antifield formalism.
This theorem states that there is no perturbatively renormalizable
non-Abelian generalization of Eqn.~(\ref{intro.alag}) with the same
field content. This is a strong result, but there are two
non-Abelian topologically massive models which evade the strictures
of this theorem because their field content are different from that
of Eqn.~(\ref{intro.naive}) even though their Abelian limits are
either Eqn.~(\ref{intro.alag}) or an equivalent first order
formulation. The modified field contents ensure that both
these models are invariant under the vector gauge transformations
of Eqn.~(\ref{intro.navsym}).

The first of these is the the Freedman-Townsend model described by
the Lagrangian
\begin{equation}
{\lag}= -\fr{1}{4}F_{\m\n}^{a}F^{a\m\n}
+\fr{m}{2}\Phi_\m^{a}\Phi^{a\m}
+\fr{m}{4}\e_{\m\n\l\s}{F(v)}^{a\m\n}B^{a\l\s}\,,
\label{intro.ftlag}
\end{equation}
where $v_\m= A_\m + \Phi_\m$, $F_{\m\n}^{a}$ is the usual
Yang-Mills field strength of $A_\mu$ and $F_{\m\n}^{a}(v)$ is the
Yang-Mills field strength calculated for $v_\mu$. The non-Abelian
2-form $B$ acts as an auxiliary field in this model, forcing $v$ to
be a flat connection.  Quantization of Freedman-Townsend model has
been studied using B-V formalism~\cite{{Leblanc:1991av}} and shown
that it is unitary but plagued with non-renormalizable
propagators. It has been shown recently that by using the
self-interaction mechanism, a first-order form of
Eqn.~(\ref{intro.alag}) gives rise to Freedman-Townsend
Lagrangian~\cite{{Khoudeir:1996mi}}.

The second one is the theory of the dynamical
2-form~\cite{Lahiri:1992hz} given by the Lagrangian
\begin{equation}
{\lag} =-\fr{1}{4}F_{\m\n}^{a} F^{a\m\n} + \fr{1}{12}
H_{\m\n\l}^{a} H^{a\m\n\l} +
\fr{m}{4}\e_{\m\n\l\s}B^{a\m\n}F^{a\l\s}, 
\label{intro.nalag}
\end{equation}
where $H_{\m\n\l}^{a}$ is now the compensated field strength,
invariant under non-Abelian vector gauge transformations,
$H_{\mu\nu\lambda}^a = \p_\m B_{\n\l}^{a} +
gf^{abc}A_\m^bB_{\n\l}^{c} +gf^{abc}C_\m^{b}F_{\n\l}^{c} + {\rm
cyclic\, terms}~$.  The quantization of this model has been studied
in the BRST/anti-BRST scheme~\cite{{Hwang:1997er},{Lahiri:1997dm}}.
A proof of renormalizability of this model was recently
constructed~\cite{{Lahiri:1999uc}} as well.  Both these models have
more fields than the na\"\i ve non-Abelian model of
Eqn.~(\ref{intro.naive}), and as a result can circumvent the no-go
theorem of~\cite{Henneaux:1997mf} and remain invariant under the
non-Abelian vector gauge transformation, given in
Eqn.~(\ref{intro.navsym}).

The purpose of this paper is threefold --- (i) to present the
Hamiltonian analysis of the na\"\i ve non-Abelian model
(\ref{intro.naive}), (ii) to see if, by any procedure one can
modify the second class constraints of the na\"\i ve non-Abelian
model to first class such that the modified theory will be
invariant under the vector gauge transformations
(\ref{intro.navsym}), and (iii) to investigate the role of the
auxiliary field in (\ref{intro.nalag}) through the analysis of
constraints. The Hamiltonian formulation of non-Abelian
topologically massive gauge theories is interesting by itself and
to the best of our knowledge has not been studied in detail.  With
these motivations, we have done the Hamiltonian analysis of the
model of Eqn.~(\ref{intro.naive}).Then we address the question
whether we can elevate, by any procedure, the na\"\i ve model to a
theory symmetric under the vector gauge transformations of
Eqn.~(\ref{intro.navsym}).

Generally a theory without any gauge symmetry can be converted to
one with a gauge symmetry by using the generalized canonical scheme
developed by Batalin, Fradkin, Tyutin
(BFT)~\cite{{Batalin:1987fm},{Batalin:1991jm}} and collaborators.
In this method, the phase space is first enlarged by introducing a
pair of canonically conjugate variables for each second class
constraint. Using these new variables, the constraints are modified
so that they have vanishing Poisson brackets among themselves. Then
the Hamiltonian is modified to have vanishing Poisson brackets with
all the modified constraints. This procedure is systematically
iterated until all second class constraints are converted to first
class and a gauge invariant Hamiltonian and a nilpotent BRST charge
are constructed.  The new fields introduced in BFT scheme are
usually identified with the St\"uckelberg fields and their momenta.
In this method one recovers the original system with second class
constraints by setting the newly introduced variables to zero. In
the path integral approach, by starting with the phase space
partition function of the embedded model one can obtain that of the
original model showing their equivalence. In order to do this, one
has to choose either the newly introduced BFT fields or
equivalently the second class constraints of the original model as
gauge fixing conditions (unitary
gauge)~\cite{{Batalin:1987fm},{Batalin:1991jm}} corresponding to
the first class constraints of the embedded model. But the
application of this method to non-Abelian theories is more involved
and there is an additional complication for the theory of
Eqn.~(\ref{intro.naive}).  As we shall argue later, BFT embedding
of the model in Eqn.~(\ref{intro.naive}) may lead to a non-local
theory.

In this paper we start from the Hamiltonian and the constraints of
Eqn.~(\ref{intro.naive}). We wish to see if the vector gauge
symmetry which was present in the Abelian model can be restored in
the form of Eqn.~(\ref{intro.navsym}) in any non-Abelian
generalization. This is done by deforming a second class
constraint, keeping the canonical Hamiltonian unchanged, so that
the modified constraint is first class and it generates the vector
symmetry transformation. This may require the modification of the
other constraints also, but the deformation is done in such a way
that the existing $SU(N)$ gauge symmetry is not lost. Finally,
given the constraints and the Hamiltonian, the Lagrangian from
which they follow has to be calculated.  This job is made easy for
the model at hand, as we already know of two non-Abelian models,
namely Freedman-Townsend model and dynamical 2-form theory, where
the vector symmetry is present.

In this procedure, the original constraints are deformed by the use
of the original phase space variables as well as new ones, which is
similar to the BFT procedure, but we look for a local theory at the
end.  Although this method as discussed seems to be applicable only
to this system at this time, it provides a clearer understanding of
the contrasting features and nature of constraints of these two
models of vector mass generation. The spirit of this method may
also be useful in dealing with other theories which lead to
non-local field theories via the BFT procedure.

The dynamical 2-form theory of Eqn.~(\ref{intro.nalag}) has an
auxiliary vector field which undergoes a transformation,
compensating for the vector gauge transformation of the
$B^a_{\mu\nu}$ field as given in Eqn.~(\ref{intro.navsym}). In this
sense, the auxiliary field resembles a St\"uckelberg
field. Typically St\"uckelberg fields are introduced to compensate
the non-invariance of a mass term. But here it is the kinetic term
of the $B^a_{\mu\nu}$ field, rather than the (topological) mass
term, which is not invariant under Eqn.~(\ref{intro.navsym}) and
requires the auxiliary field in order to restore invariance.
Another difference between a St\"uckelberg field and the auxiliary
field is that a St\"uckelberg field generally has a kinetic term
while the auxiliary field here does not have any.  Thus in the role
of a compensating field, the auxiliary field appears just as in the
St\"uckelberg mechanism, but its properties are very different from
those of a St\"uckelberg field. In the course of our Hamiltonian
analysis of the system, we shall investigate the role of the
auxiliary field in the model of Eqn.~(\ref{intro.nalag}) in the
Hamiltonian formulation.

This paper is organized as follows. In Sec.~{\ref{fof}}, we present
the canonical analysis of a first-order formulation of the theory
of Eqn.~(\ref{intro.naive}) and show that it has first class as
well as second class constraints. Using the constraints and their
Poisson brackets we then argue that BFT embedding can lead to
non-local theory.  In Sec.~{\ref{F-T}), we deform the constraints
using only the original phase space variables so that the vector
gauge symmetry is implemented by first class constraints. There we
show that the deformed system is equivalent to the
Freedman-Townsend model.  Next in Sec.~{\ref{sof}}, we deform the
constraints using the original as well as newly introduced
canonically conjugate variables, whereby we obtain the dynamical
2-form theory. In Sec~{\ref{secf}}, we discuss the role played by
the auxiliary $C_\mu$ field in dynamical 2-form theory of
Eqn.~(\ref{intro.nalag}). We also show here that the model in
Eqn.~(\ref{intro.nalag}) cannot be the embedded version of the
na\"\i ve model of Eqn.~({\ref{intro.naive}). We conclude with a
comparative study of the Freedman-Townsend model and dynamical
2-form theory and discussions in Sec.~{\ref{disc}}.

{\em Conventions:} We use the metric $g_{\m\n} = {\rm diag}(1, -1,
-1, -1)$ and $\e_{0123} = 1$.  We shall take the gauge group to be
$SU(N)$, with generators $t^a$ satisfying
\begin{eqnarray}
\left[ {t}^a,{t}^b\right] &=& if^{abc}{t}^c\,,\nonumber \\
tr ({t}^a{t}^b) &=& \fr{1}{2}\d^{ab}\,.
\end{eqnarray}
The field strength $F$ of a gauge field $A$ is defined as
\begin{equation}
F_{\m\n}^a =\p_\m A_\n^{a}-\p_\n A_\m^{a} +
gf^{abc}A_\m^{b}A_\n^{c}\,. 
\end{equation}
%

\section{Na\" ive Non-Abelian Model: Hamiltonian Analysis}\label{fof}
%
In this section we consider Eqn.~(\ref{intro.naive}) in a
first-order formulation. We analyze the Hamiltonian structure of
the system. The non-invariance of the model under the vector gauge
transformation of Eqn.~(\ref{intro.navsym}) is reflected by the
second class nature of the constraint which implements this
symmetry. By deforming this constraint using only the original
phase space variables we can convert this constraint to a first
class one. This requires modification of the remaining constraints
as well so as to leave the first class or second class nature of
all the other constraints unchanged. Thus this deformation gives us
a new gauge system which is invariant under the $SU(N)$ gauge
symmetry of Eqn.~(\ref{intro.gaugesym}), as well as the vector gauge
symmetry of Eqn.~(\ref{intro.navsym}). We shall find that the new
system is identical to the Freedman-Townsend model.
 
We start the Hamiltonian analysis from the first-order Lagrangian
\begin{eqnarray}
{\lag}= -\fr{1}{4}F_{\m\n}^{a}F^{a\m\n} +\fr{m^2}{2} \Phi_\m^{a}
\Phi^{a\m} +\fr{m}{2}\e_{\m\n\l\s}\Phi^{a\m}(D^\n B^{\l\s})^a \no\\
+ \fr{m}{4}\e_{\m\n\l\s}A^{a\m}(D^\n B^{\l\s})^a
+\fr{m}{4}\e_{\m\n\l\s}A^{a\m}\p^\n B^{a\l\s}.
\label{fof.lag}
\end{eqnarray}
By integrating out the $\Phi_{\mu}^a$ field from the above
Lagrangian we get the second-order Lagrangian of
Eqn.~(\ref{intro.naive}). We have not put the gauge potential part
in first order form. The primary constraints following from this
Lagrangian are
\begin{eqnarray}
\Pi_{0}^a &\we& 0\,,\quad {\t \Pi}_{0}^a\we 0\,,\quad\Pi_{0i}^a\we 0\,,
\quad{\t \Pi}_{i}^a\we 0\,,\nonumber \\
\L_{ij}^a &=& \Pi_{ij}^a+m\e_{ijk}(A^{ak} + \Phi^{ak})\we0\,,
\label{fof.primary}
\end{eqnarray}
where $\epsilon_{ijk} = \epsilon_{0ijk}$, and we have written
$\Pi^a_\mu, \tilde\Pi^a_\mu, \Pi^a_{\mu\nu}$ for the momenta
canonically conjugate to $A^{a\mu}, \Phi^{a\mu}$ and $B^{a\mu\nu}$,
respectively. The total Hamiltonian is then
\begin{eqnarray}
{\ham}_T = \fr{1}{4}{F}_{ij}^{a}F^{aij} -\fr{1}{2}{\Pi}_i^{a} \Pi^{ai}
-\fr{m^2}{2}{\Phi}_i^{a}\Phi^{ai}
- A_0^a\tilde\Lambda^a -
\Phi_0^a\tilde\omega^a - B^{a0i}\tilde\Lambda^a_i \,.
\label{fof.canham}
\end{eqnarray}

The coefficients of the Lagrange multipliers $A^{a0},
\Phi^{a0},B^{a0i}$ are the secondary constraints which appear upon
demanding the persistence of the primary constraints, 
\begin{eqnarray}
{\t \L}^a &=& \PB{\Pi_0^a}{\ham} = (D^i\Pi_i)^a
+\fr{m}{2}\eijk(D^i B^{jk})^a +\fr{m}{2}
gf^{abc}\eijk\Phi^{bi}B^{cjk}\,,
\label{fof.tLa} \\ 
\tilde\Lambda_i^a &=& \PB{\Pi_{0i}^a}{\ham} =
\fr{m}{2}\eijk(\Phi^{ajk}+F^{ajk})\,,
\label{fof.tLia} \\ 
\tilde\omega^a &=& \PB{\tilde\Pi_0^a}{\ham} = m^2 \Phi^{a0}
+\fr{m}{2}\eijk(D^i B^{jk})^a\,.
\label{fof.tomega}
\end{eqnarray}
All these secondary constraints are second class as can be checked
easily from their Poisson brackets with one another. But we can
construct their linear combinations which are first class at this
stage. Let us define one combination which is going to be the Gauss
law constraint, and another which will generate the vector gauge
transformations,
\begin{eqnarray}
\Lambda^a &=& \tilde\Lambda^a -\half gf^{abc}\L_{ij}^b
B^{cij} -gf^{abc} {\t \Pi}_{0}^b\Phi^{c0}
-gf^{abc} {\t \Pi}_{i}^b\Phi^{ci} \,,
\label{fof.La}\\
{\bar\L}_{i}^a &=& \tilde\Lambda_i^a - (D^j\L_{ij})^a \,.
\label{fof.barLia}
\end{eqnarray}
At this stage, the Poisson brackets among the constraints are
\begin{eqnarray}
\PB{\L^a}{\L^b} &=&
-gf^{abc} \L^c\,,\quad\PB{ \L^a}{{\bar\L}_i^{b}} =
gf^{abc} \L_i^{c}\,, 
\quad 
\PB{\L^a}{\tilde\w^b} = gf^{abc}{\tilde\w}^c\,,\nonumber \\ 
 \PB{ \L^a}{\L_{ij}^b} &=& gf^{abc}\L_{ij}^c\,, \quad
\PB{\L^a}{{\t\Pi}_{i}^{a}} = gf^{abc}{\t \Pi}_{i}^c\,,\quad
\PB{\L^a}{\Pi_{0}^b} = 0\,,\quad 
\PB{\L^a}{\Pi_{0i}^b} = 0\,,
\label{pb.pb1}\\
\nonumber \\
\nonumber \\
\PB{{\bar \L}_{i}^a}{{\bar\L}_{j}^b} &=& 0 \,,\quad
\PB{{\bar\L}_{i}^a} {{\tilde\w}^b} = 0 \,, \quad 
\PB{{\bar\L}_{i}^a} {\Pi_{0}^b} = 0\,,
\quad   
\PB{{\bar\L}_{i}^a} {{\t \Pi}_{0}^b} = 0 \,, \quad
\PB{{\bar\L}_{i}^a}{{\t \Pi}_{j}^b} = 0 \,,\nonumber\\
\PB{{\bar\L}_{i}^a}{\L_{ij}^b} &=& 0\,, \quad 
\PB{{\t \Pi}_{0}^a}{\tilde\w^b}= -m^2 \,,\nonumber\\
\PB{{\t \Pi}_{i}^a}{\L_{ij}^b} &=& -m\delta^{ab}\eijk \,\qquad
\PB{{\t\w}^a}{\L_{ij}^b}=m\e_{ijk}D^{abk}\d(x-y).
~
\label{pb.pb2}
\end{eqnarray}
Note that both $\Lambda^a$ and $\bar\Lambda^a_i$ have vanishing
Poisson brackets with all the other constraints as well as with
themselves. The constraint $\Lambda^a$ is preserved,
$\PB{\Lambda^a}{\ham} = 0$, while
\begin{eqnarray}
\PB{{\bar \L}_{i}^a}{\ham} = gf^{abc}\Pi^{bj}\L_{ij}^c 
-m g f^{abc}\eijk\Pi^{bj}\Phi^{ck} \equiv \psi_i^a
\label{fof.ter}
\end{eqnarray}
is a tertiary constraint, which has non vanishing Poisson bracket
with ${\bar\Lambda}_{i}^a$ and ${\t\w}^a$.  Now the first class
constraints are $\Pi_0^{a}, \Pi_{0i}^a$ and $\L^a$, and the second
class constraints are ${\t\w}_i^{a}, \L_{ij}^{a}, \w^a, {\t
\Pi}_0^{a}, \bar\Lambda_i^{a},~\psi_i^{a}$ which together remove 24
out of 28 phase space degrees. Thus a na\"\i ve counting shows that
the above model has only four phase space degrees of freedom and
cannot describe a massive spin-one theory.  On the other hand, the
free part of the action (i.e., the $g \to 0$ limit) coincides with
the Abelian action, which has six phase space degrees of freedom
and in fact describes a massive vector field. It is therefore
possible that there is a reducibility among second class
constraints which is not manifest, in which case the model
may still describe massive spin-one particles.

The second class nature of the constraint $\bar\L_{i}^a$ is
expected since the model of Eqn.~(\ref{intro.naive}), and therefore
Eqn.~(\ref{fof.lag}), does not have the vector gauge invariance of
Eqn.~(\ref{intro.navsym}).  Generally, by applying the BFT
procedure, a theory with second class constraints $T_{\a}$ (with
matrix of Poisson brackets $\{{T_\a},{T_\b}\}=\D_{\a\b}$) and
Hamiltonian can be converted to a theory with only first class
constraints and gauge invariant Hamiltonian. In this method one
first enlarges the phase space by introducing auxiliary variables
$\th_\a$ corresponding to each of the second class constraints. 
These variables satisfy
\begin{eqnarray}
\{\th_\a,\th_\b\} = \w_{\a\b},
\label{bft.pb}
\end{eqnarray}
which may be taken to be field-independent, and $\w_{\a\b}$ is such
that ${\rm det}|\w_{\a\b}|\ne0$.  Now we define the first class
constraints ${\bar T}_{\a}(P, Q, \th_\a)~$ (where $P$ and $Q$ stand
for the original canonically conjugate phase space variables) in
the extended phase space, satisfying
\begin{eqnarray}
\{{\bar T}_\a, {\bar T}_\b\}=0\,.
\label{fpb}
\end{eqnarray}
The solution for this is obtained in a series form as
\begin{eqnarray}
{\bar T}_\a= T_{\a} + X_{\a\b}\th^\b +{\rm higher~order~terms~in~
}\th_\a \,, 
\label{sol.pb}
\end{eqnarray}
where $X_{\a\b}$ satisfy
\begin{eqnarray}
X_{\a\b}\w^{\b\l} X_{\l\r}=\D_{\a\r}.
\label{bft.cond}
\end{eqnarray}

After converting the second class constraint to strongly involutive ones,
one proceeds to construct the gauge invariant Hamiltonian ${\bar
H}(P,Q,\th_\a)$ in the extended phase space. This gauge invariant
Hamiltonian has to satisfy
\begin{eqnarray}
\{{\bar T}_\a, {\bar H}\}=0.
\label{bft.cod1}
\end{eqnarray}
Solving the above equation gives ${\bar H}$ in a series form.

The first class constraints and gauge invariant Hamiltonian are
calculated by solving Eqs.~(\ref{sol.pb}) and (\ref{bft.cod1}) in
BFT procedure by iteration.  However, to solve Eqs.~(\ref{sol.pb})
and (\ref{bft.cod1}) one needs the inverse of both $\w_{\a\b}$ and
$X_{\a\b}$. In the case of the na\"\i ve model, it can be seen
directly from Eqn.~(\ref{pb.pb2}) that the matrix $\D_{\a\b}$ of
the Poisson brackets between the second class constraints $(T_\a)$
involve derivatives of delta functions.  Because of this $X_{\a\b}$
will also have derivatives of delta functions (see
Eqn.~(\ref{bft.cond})) and therefore its inverse is likely to be
non-local. This will result in the non-locality of the first class
constraints ${\bar T}_\a$ and gauge invariant Hamiltonian ${\bar
H}$. Thus from the constraint structure of the na\"\i ve
non-Abelian model, we see that the usual BFT embedding will very
likely lead to a non-local theory.

For the sake of completeness we should mention that 
there is an alternate procedure of converting theories with second
class constraints to theories with only first class constraints
known as the gauge unfixing method~\cite{{Mitra:1990mp}}. In all
known examples, this procedure and BFT embedding result in the same
first class theory. In Appendix~\ref{a2}, we apply the gauge
unfixing procedure to na\"\i ve non-Abelian model of
Eqn.~(\ref{intro.naive}), and find that it too fails to give a
first-class theory.

\section{Freedman-Townsend Model}\label{F-T}
%
Since we want the modified theory to be local as well as invariant
under the vector gauge transformation as well as under all the original
symmetry transformations of the 
model, in this section we adopt a different approach to convert the
second class constraints. Here we would like to modify the
constraints $~{\bar\L}_i^{a}\rightarrow{\L_i^{a}}~$ such that
$\PB{\L_i^{a}} {\ham}\we 0$ as well as $\PB{\L_{i}^a} {\chi}\we0$,
for all constraints $\chi$ in the theory. Since it is only
$\psi_i^{a}$ which has a non-vanishing Poisson bracket with ${\bar
\L}_{i}^a$, we start by modifying ${\bar \L}_{i}^a$ such that
$\PB{\bar\Lambda_i^a}{\psi_j^b} \approx 0$. But this modification
will change the other Poisson bracket relations in
Eqs.~(\ref{pb.pb1}) and (\ref{pb.pb2}). In order to keep those
initial first class constraints as first class and also to have
$\L_i^{a}$ first class, we have to further modify $\L_i^{a}$ as
well as other constraints. For the sake of convenience, let us
define $v_{i}^a=A_{i}^a+\Phi_{i}^a$ and
$D(v)^{abi}=\d^{ab}\p^i+gf^{acb}v^{ci}.$ Then the modified
constraints read
\begin{eqnarray}
\L^a 
&=& (D(A)^i\Pi_i)^a + \fr{m}{2}\eijk(D(v)^aB^{jk})^a-
\fr{g}{2}f^{abc}\L_{ij}^bB^{cij}-gf^{abc}{\t
\Pi}_{0}^b\Phi^{c0}-gf^{abc}{\t \Pi}_{i}^b\Phi^{ci}\,,
\label{fof.Lafinal}\\
\L_{i}^a&=& \bar\Lambda_i^a
+\fr{m}{2}\e_{ijk}f^{abc}\Phi^{bj}\Phi^{ck}-gf^{abc}\Phi^{bj}
\L_{ij}^c  
\label{fof.Liafinal}\\
\w^a&=& \tilde\omega^a +\fr{m}{2}gf^{abc}\e_{ijk} \Phi^{bi}B^{cjk} 
\nonumber \\ 
&=& m^2\Phi_{0}^a +\fr{m}{2}\eijk(D(v)^i B^{jk})^a\,.
\label{fof.wfinal}
\end{eqnarray}
The remaining constraints of Eqn.~(\ref{fof.primary}) are unchanged.
The algebra of the modified constraints is
\begin{eqnarray}
\{\L^a, \L^b\}&=&-gf^{abc}\L^c,~~~\{\L^a, \L_{i}^b\} =
gf^{abc}\L_{i}^c,\no\\ 
\{\Pi_{0}^a,\w^b\}&=&-m^2,~~\{\L_{i}^a,\L_{j}^b\}=0\,,\no\\
\{\L^a,\w^b\} &=& gf^{abc}\w^c,\no\\
\{\L_{i}^a, \w^b\}&=& \fr{g}{2}f^{abc}\eijk F(v)^{cjk}\,.
\label{alg.lab}
\end{eqnarray}
Note that the last Poisson bracket in the above appears to be still
non-vanishing (even weakly), but is not so. This can be seen by
noting that ${\L}_{i}^a$ can be expressed as 
\begin{eqnarray}
\L_{i}^a=
\fr{m}{2}\e_{ijk} F(v)^{ajk} - (D(v)^j\L_{ij})^a,
\label{newl}
\end{eqnarray}
We want the modified theory to be of first-order in $B^a_{\mu\nu}$,
and consequently $\L_{ij}^a \approx 0$ should remain a constraint.
Therefore the first term in $\L_{i}^a$ viz.,
${\displaystyle\fr{m}{2}\e_{ijk} F(v)^{ajk}}$ must be constrained
to vanish by itself.  Hence the last Poisson bracket in
Eqn.~(\ref{alg.lab}) vanishes weakly. Thus we see that the
constraint $\L_{i}^a$ is first class. {}From (\ref{alg.lab}) it is
clear that $\L^a$ is first class and $\w^a$ is second class.

Also we notice that
\begin{eqnarray}
(D(v)^i\L_i)^a=-\fr{g}{2}f^{abc} F(v)^{bij}\L_{ij}^c\,,
\end{eqnarray}
which is zero upon using the constraint ${\displaystyle
\fr{m}{2}\eijk F(v)^{ajk}} \approx 0.$ Thus we see that $\L_{i}^a$
is reducible on the constraint surface. Thus among the first class
constraints $\Pi_{0}^a,~\Pi_{0i}^a,~\L^a,~\L_{i}^a$ only seven are
linearly independent, and we also have eight second class
constraints ${\t\Pi}_{0}^a,\, {\t\Pi}_{i}^a,\, \L_{ij}^a,\, \w^a$.
Because of these constraints, six phase space degrees of freedom
remain, and the system defined by these constraints
(\ref{fof.primary}, \ref{fof.Lafinal}, \ref{fof.Liafinal},
\ref{fof.wfinal}) and the canonical Hamiltonian
\begin{eqnarray}
{\ham}_c = \fr{1}{4}{F}_{ij}^{a}F^{aij}
-\fr{1}{2}{\Pi}_i^{a} \Pi^{ai} -\fr{m^2}{2}{\Phi}_i^{a}\Phi^{ai}\,
\label{ham1.final}
\end{eqnarray}
describes a massive spin-one model.

The covariant Lagrangian from which this set of constraints and
Hamiltonian follow is
\begin{eqnarray}
{\lag}=-\fr{1}{4}F_{\m\n}^aF^{a\m\n} +\fr{m^2}{2}\Phi_{\m}^a\Phi^{a\m} 
+\fr{m}{4}\e_{\m\n\l\s}v^{a\m}(D(v)^\n B^{\l\s})^a 
+\fr{m}{4}\e_{\m\n\l\s}v^{a\m}\p^\n B^{a\l\s}.
\end{eqnarray}
This Lagrangian can be rewritten after an integration by
parts as 
\begin{eqnarray}
{\lag}=-\fr{1}{4}F_{\m\n}^aF^{a\m\n} +\fr{m^2}{2}\Phi_{\m}^a\Phi^{a\m} 
+\fr{m}{4}\e_{\m\n\l\s}F(v)^{a\m\n} B^{a\l\s}, 
\end{eqnarray}
which is the Freedman-Townsend Lagrangian describing a massive
spin-one theory.

\section{Dynamical 2-form Theory}\label{sof}
%
In this section we introduce new pairs of canonically conjugate
variables and deform the constraints using these as well as the
original phase space variables. The goal of this deformation is
again to turn the constraint of Eqn.~(\ref{fof.tLia}) into a first
class constraint. Here again we start our Hamiltonian analysis from
a first order Lagrangian
\begin{eqnarray}
{\lag} = -\fr{1}{4}F_{\m\n}^{a}F^{a\m\n} +\fr{m^2}{2} \Phi_\m^{a}
\Phi^{a\m} +\fr{m}{4}\e_{\m\n\l\s}\Phi^{a\m\n}B^{a\l\s} \no\\ +
\fr{m}{4}\e_{\m\n\l\s}A^{a\m}(D^\n B^{\l\s})^a
+\fr{m}{4}\e_{\m\n\l\s}A^{a\m}\p^\n B^{a\l\s},
\label{sof.lag}
\end{eqnarray}
where $\Phi_{\m\n}^{a}=(D_\m\Phi_\n-D_\n\Phi_\m)^{a}$. This
Lagrangian looks different from the Lagrangian of
Eqn.~(\ref{fof.lag}), but the difference is only by a total
derivative. By eliminating $\Phi_\m^{a}$ we will again get back
the second-order Lagrangian of Eqn.~(\ref{intro.nalag}).

The structure of the constraints will be different, however. The
primary constraints following from this Lagrangian are
\begin{eqnarray}
{\Pi}_0^{a}\we0,~~{\Pi}_{0i}^{a}\we0,~~{{\t \Pi}}_0^{a}\we0, \no\\
{\t\w}_i^{a} = {\t \Pi}_i^{a} -\fr{m}{2}\eijk B^{ajk}\we0,\no\\
{\L}_{ij}^{a}= {\Pi}_{ij}^{a} +m\e_{oijk}A^{ak}\we0,
\label{prim.dyn}
\end{eqnarray}
where $\Pi_{\m}^a,~\Pi_{\m\n}^a,~{\t\Pi}_{\m}^a$ are the momentum
conjugates of $A^{a\m},~B^{a\m\n},~ \Phi^{a\m}$ respectively.  The
total Hamiltonian is then
\begin{eqnarray}
{\ham}_T &=& \fr{1}{4}{F}_{ij}^{a}F^{aij} -\fr{1}{2}{\Pi}_i^{a}
\Pi^{ai} -\fr{m^2}{2}{\Phi}_i^{a}\Phi^{ai}\, \nonumber\\ 
&& \qquad\qquad - A_0^a\tilde\Lambda^a -
\Phi_0^a\tilde\omega^a - B^{a0i}\tilde\Lambda^a_i \,,
\label{sof.canham}
\end{eqnarray}
where $\tilde\Lambda^a, \tilde\omega^a$ and $\tilde\Lambda^a_i$ are
the secondary constraints and are the same as the constraints
denoted by the same symbols in
Eqn.~(\ref{fof.tLa}-\ref{fof.tomega}), each of which have
non-vanishing Poisson bracket with at least one of the remaining
constraints.  We define the linear combinations
\begin{eqnarray}
\L^a&=& {\t\L}^a - \fr{g}{2}f^{abc}{\L}_{ij}^{b}B^{cij} -gf^{abc}
{{\t \Pi}}_0^{b} \Phi^{c0}-gf^{abc}{\t\w}_i^{b}{\Phi}^{ci},\\
{\bar\L}_i^{a}&=&{{\t\L}}_i^{a} -(D^j\L_{ij})^a,
\end{eqnarray}
The constraints $\Pi_0^{a},~\Pi_{0i}^{a},~\L^a, ~{\rm
and}~{\bar\L}_i^{a}$ have vanishing (at least weakly) Poisson
brackets with all constraints. The Poisson bracket of $\L^a$ with
the canonical Hamiltonian vanishes weakly but that of
${\bar\L}_i^{a}$ as in the previous section gives a tertiary
constraint
\begin{eqnarray}
\psi_i^{a} = \PB{{\bar\L}_i^a}{\ham}
= gf^{abc}\Pi^{bj}\L_{ij}^{c}-m\eijk gf^{abc}\Pi^{bj} \Phi^{ck},
\end{eqnarray}
which has non-vanishing Poisson bracket with ${\bar\L}_i^{a}$.  The
Poisson brackets among the constraints are same as in
Eqs.~(\ref{pb.pb1}) and (\ref{pb.pb2}). Since here also we see that
BFT embedding will lead only to a non-local theory, we adopt an
alternate approach to modify the constraints. 

As in the previous section, first we modify ${\bar\L}_i^{a}$ to
$\L_{i}^a$ such that $\PB{\L_{i}^a}{\ham} \we0.$ As before, this
modification of ${\bar\L}_{i}^a$ changes its Poisson brackets of
with all other constraints.  So we further modify $\L^a,~{\bar
\L}_{i}^a,~\w^a$ and $\t\w_{i}^a$ such that the modified
constraints $\L^a,~\L_{i}^a$ are in involution with all
constraints. Unlike in the earlier section here we modify the
constraints by introducing canonically conjugate pairs
$(C_{i}^a,~P^{bj})$, as in the BFT formalism.  Thus we get the
following modified first class constraints
\begin{eqnarray}
{\t \w}_{i}^a &=& {\t \Pi}_{i}^a -\fr{m}{2}\eijk
(B^{ajk}-C^{ajk})\,, \\
\L^a &=& {\t\L}^a - \fr{g}{2}f^{abc}{\L}_{ij}^{b}B^{cij} -gf^{abc}
{{\t \Pi}}_0^{b} \Phi^{c0} - gf^{abc}{\t\w}_i^{b}{\Phi}^{ci}
- gf^{abc}\chi_{i}^bC^{ci} \label{sof.lamfin}\\
{\L}_i^{a} &=& {{\t\L}}_i^{a} -(D^j\L_{ij})^a -
\chi_{i}^a,\label{sof.Lambdafinal}\\ 
\w^a &=& m^2 \Phi_{0}^a +\fr{m}{2}\eijk D^{abi}(B^{jk}-C^{jk})^b \,.
\end{eqnarray}
where $C_{ij}^a= (D_i C_j -D_j C_i)^a$ and $ \chi_{i}^a = P_{i}^a 
+ \displaystyle{\fr{m}{2}\eijk\Phi^{ajk}}.$ 
At this stage we have 8 first class and 11 second class constraints. 

Here we  note that the combination which is left invariant
by the first class constraints of Eqn.~(\ref{sof.Lambdafinal}) is
\begin{eqnarray}
B_{ij}^a-(D_i C_j -D_j C_i)^a
\label{sof.vector}
\end{eqnarray}
with the vector gauge transformations given by 
\begin{eqnarray}
\d(B_{ij}^a) &=& (D_i \l_j-D_j\l_i)^a,\nonumber\\
\d(C_{i}^a) &=& \l_{i}^a.
\label{symred}
\end{eqnarray}
Obviously, the combination of Eqn.~(\ref{sof.vector}) has a
further invariance, under
\begin{eqnarray}
\d(C_{i}^a)=(D_i \theta)^a,~~~\d (B_{ij}^a)=gf^{abc}F_{ij}^b \theta^c.
\label{com.sym}
\end{eqnarray}
Because of this invariance in Eqn.~(\ref{com.sym}), the gauge
transformation generated by $\L_{i}^a$ are not mutually
independent.  Since these reducible transformations of
Eqs.~(\ref{symred}) and (\ref{com.sym}) are gauge symmetries of the
theory, the constraints that generate these transformations must be
first class constraints.  So we first enlarge the phase space by
introducing another pair of conjugate variables $C_{0}^a$ and
$P^{a0}$ and demand that (i)~$P^{a0}$ is a primary constraint and
(ii)~the total Hamiltonian of the modified theory should contain a
term $-C_{0}^a\T^a$ so that $\T^a$ is a secondary constraint. The
form of $\T^a$ is such that it generates the transformation
(\ref{com.sym}) and it is linearly dependent on $\L_{i}^a$.  This
fixes the form $\T^a$ to be
\begin{eqnarray}
\T^a = \fr{m}{2}\eijk(D^i\Phi^{jk})^a -\fr{1}{2}gf^{abc}F^{bij}\L_{ij}^c
-(D^i\chi_i)^a \equiv -(D^i\L_i)^a.
\end{eqnarray}
which makes the generator of the transformation of
Eqn.~(\ref{symred}) reducible. $\T^a$ has vanishing Poisson
brackets with all other constraints and $\{\T^a, {\ham}\}$ gives a
tertiary constraint,
\begin{eqnarray}
\PB{\Theta^a}{\ham} = \s^a = mgf^{abc}\eijk(D^i\Pi^j)^b\Phi^{ck}
-gf^{abc}\Pi^{bi}{\w^c}_i -gf^{abc}(D^i\Pi^j)^d{\L}_{ij}^{c}
\nonumber\\ -mgf^{dbc}\eijk D^{adi}(\Pi^{bj}\Phi^{ck}) -
m\fr{g}{2}f^{abc}\eijk F^{bij}\Pi^{ck}.
\end{eqnarray}
%
The Poisson brackets of $\s^a$ with $\L^a$ and $\L_i^{a}$ vanish
weakly and that with $\T^a$ strongly. There are no further
constraints as $\s^a$ has non-zero Poisson bracket with
${{\t\w}}_i^{a}$ which makes it second-class. Thus we have obtained
all the constraints of the modified theory.

Thus now the expanded theory has the first class constraints
\begin{eqnarray}
\Pi_{0}^a,~\Pi_{0i}^a,~P_{0}^a,~\L^a, \L_{i}^a,~\T^a,
\end{eqnarray}
and second class constraints 
\begin{eqnarray}
{\t \w}_{i}^a,~\w_{i}^a,~\L_{ij}^a,~\w^a,~{\t \Pi}_{0}^a,~\s^a.
\end{eqnarray}
The canonical Hamiltonian remains
\begin{eqnarray}
{\ham}_c = \fr{1}{4}{F}_{ij}^{a}F^{aij} -\fr{1}{2}{\Pi}_i^{a}
\Pi^{ai} -\fr{m^2}{2}{\Phi}_i^{a}\Phi^{ai}. 
\label{ham2.final}
\end{eqnarray}
The 9 linearly independent first class constraints along with 12
second class ones will leave 6 phase space degrees of freedom, and
thus the model with the above constraints and ${\ham}$ describes a
massive spin-one theory.

The Lagrangian from which this constraints and ${\ham}$ follow is
\begin{eqnarray}
{\lag}= -\fr{1}{4}F_{\m\n}^{a}F^{a\m\n} +\fr{m^2}{2} \Phi_\m^{a} \Phi^{a\m}
+\fr{m}{4}\e_{\m\n\l\s}\Phi^{a\m\n}B^{a\l\s} -\fr{m}{4}\e_{\m\n\l\s}
C^{a\m\n}\Phi^{a\l\s} \nonumber\\
+ \fr{m}{4}\e_{\m\n\l\s}A^{a\m}(D^\n B^{\l\s})^a
+\fr{m}{4}\e_{\m\n\l\s}A^{a\m}\p^\n B^{a\l\s},
\end{eqnarray}
where $C_{\m\n}^{a}=(D_\m C_\n-D_\n C_\m)^{a}$. 
By eliminating $\Phi_\m^{a}$ using its equation of motion, we
get a second-order Lagrangian, which up to a total derivative is
\begin{eqnarray}
{\lag} =-\fr{1}{4}F_{\m\n}^{a} F^{a\m\n} + \fr{1}{12} H_{\m\n\l}^{a}
H^{a\m\n\l} +\fr{m}{4}\e_{\m\n\l\s}B^{a\m\n}F^{a\l\s},
\end{eqnarray}
where  $H_{\m\n\l}^{a} = \p_\m B_{\n\l}^{a} + gf^{abc}A_\m^bB_{\n\l}^{c}
+gf^{abc}C_\m^{b}F_{\n\l}^{c} + cyclic~terms$.

\section{ Role of the auxiliary Field $C_\mu$}\label{secf}

In this section we analyze the role of $C_\mu$ field in the
dynamical 2-form theory of Eqn.~(\ref{intro.nalag}). Na\"\i vely,
because of its gauge transformation property, $C_\mu$ seems to be a
St\"uckelberg field in Eqn.~(\ref{intro.nalag}) for the na\"\i ve
model of Eqn.~(\ref{intro.naive}), compensating for the vector
gauge symmetry of Eqn.~(\ref{intro.navsym}).  Usually the
St\"uckelberg field is a dynamical field introduced to compensate
for the non-invariance of the mass term under local gauge
transformations.  On the other hand, here the (topological) mass
term in Eqn.~(\ref{intro.naive}) is invariant under the vector
gauge transformation whereas the kinetic term is not. The
invariance of kinetic term under Eqn.~(\ref{intro.navsym}) is
restored by the compensating transformation of the $C_\mu$
field. Also here the $C_\mu$ field is non-dynamical, as no kinetic
term appears for it in the Lagrangian. The question that arises
naturally is therefore --- what is the nature of the field $C_\mu$?

One way of understanding the role of $C_\mu$ field is to ask if the
dynamical 2-form theory Eqn.~(\ref{intro.nalag}) is an embedded
version of the na\"\i ve non-Abelian model of
Eqn.~(\ref{intro.naive}) obtained by converting second class
constraints $\bar\Lambda_i^a$ to first class constraints $\L_{i}^a$
as in Eqn.~(\ref{sof.Lambdafinal}), which generate the vector
symmetry of Eqn.~(\ref{intro.navsym}) in the extended model of
Eqn.~(\ref{intro.nalag}). The partition function of an embedded
gauge theory is known to reduce to that of the original model with
the choice of the second class constraints $(T_\a)$ of the original
model as gauge fixing conditions for the first class constraints
$({\bar T}_\a)$ of the embedded model. Equivalently one can also
choose the newly introduced BFT variables (here these are $C_{0}^a,
P_{0}^a, C_{i}^a\ {\rm and}\ P_{i}^a)$ as the gauge fixing
conditions~\cite{{Batalin:1987fm},{Batalin:1991jm}}.

Consider the phase space partition function of the
dynamical 2-form theory
\begin{eqnarray}
Z= \int {\cal D}\eta\,\d(\chi^\a)\,\d(F^\a)\,\d(G^\b)~\Delta_{FP}
\det\{\chi^\a,\chi^\b\}\,\exp^{\int d^4 x (P\p^0 Q-H)} \,,
\label{part}
\end{eqnarray}
where the measure ${\cal D}\eta$ run over all phase space
variables, $\chi^\a$ are the second class constraints, $F^\a$ are
the first class constraints, $G^\b$ are the corresponding gauge
fixing conditions and $\Delta_{FP}$ is the Faddeev-Popov determinant
of the embedded model.  Here $P$ and $Q$ stand for the generic
momenta and fields. Let us suppose that the dynamical 2-form
theory is the embedded version of the na\"\i ve non-Abelian
model. Then if we choose the second class constraints of the
na\"\i ve model (either ${\bar \L}_{i}^b$ or $ \psi_{i}^b$ of
Eqs.~(\ref{fof.barLia}), (\ref{fof.ter}) respectively) as gauge
fixing conditions corresponding to the first class constraints
$\L_{i}^a$ of Eqn.~(\ref{sof.Lambdafinal}), the partition function
(\ref{part}) must reduce to that of na\"\i ve non-Abelian
model. For the other first class constraints of dynamical 2-form
theory
\begin{eqnarray}
F^\a =
(\Pi_{0}^a,~\Pi_{0i}^a,~P_{0}^a,~\L^a), 
\end{eqnarray}
we choose the gauge fixing conditions as
\begin{eqnarray}
G^\a= (A^{b0},~B^{b0l},~C^{b0},~\p^i A_{i}^b)
\end{eqnarray}
respectively.  With the above choice of gauge fixing conditions
(either with choice ${\bar \L}_{j}^b$ or $\psi^{bj}$ as the gauge
conditions for $\L_{i}^a$), it is easy to see that the
Faddeev-Popov determinant vanishes. Thus the partition
function does not reduce to that of na\"\i ve non-Abelian model.

Instead of choosing the original second class constraints as gauge
fixing conditions, equivalently one could choose the newly
introduced BFT variables as the gauge fixing conditions. Here as we
have seen, $P_{0}^a$ by itself is a first class constraint (it is
not appearing in any of the modified constraints) and we have
chosen $C^{b0}$ as its gauge fixing condition. We chose $C^{bj}$ as
the gauge condition for $\L_{i}^a$ of Eqn.~(\ref{sof.Lambdafinal})
since their Poisson bracket is non-vanishing. But we cannot choose
$P_{i}^a$ as part of a gauge-fixing condition for any of the first
class constraints of the dynamical 2-form theory.  Consequently, we
cannot implement the vanishing condition of all newly introduced
BFT variables as gauge conditions and get back the original model
(na\"\i ve non-Abelian model).

Thus we see that even when we choose the unitary gauge condition,
the partition function of dynamical 2-form theory Eqn.~(\ref{part})
does not reduce to that of the na\"\i ve non-Abelian model. Hence
the dynamical 2-form theory of Eqn.~(\ref{intro.nalag}) and the
na\"\i ve non-Abelian model of Eqn.~(\ref{intro.naive}) are not
simply related by BFT embedding.  In the covariant quantization
scheme, this can be seen from the fact that implementing $C_\mu=0$
as a gauge fixing condition for the vector symmetry
(\ref{intro.navsym}) is not proper as the quadratic part of 2-form
B will still be non-invertible.

\section{Conclusion}\label{disc}
In this paper, we have studied three different non-Abelian
generalizations of the topologically massive Abelian gauge theory
in $3+1$ dimensions, given in Eqn.~(\ref{intro.alag}). In
Sec.~\ref{fof}, using the canonical analysis of the na\"\i ve
non-Abelian model of Eqn.~(\ref{intro.naive}), which is not
invariant under the vector gauge transformations of
Eqn.~(\ref{intro.navsym}), we have shown that the BFT Hamiltonian
embedding of this model will lead to a non-local theory. Then in
Sec.~\ref{F-T}, starting with the na\"\i ve non-Abelian model and
using an alternate approach we have shown that by a deformation of
the constraints we can obtain the Freedman-Townsend model. Here we
have used only the original phase space variables to modify the
constraints. We have also shown the off-shell reducibility of the
latter model. In Sec.~\ref{sof}, by a different modification of the
constraints of na\"\i ve model, where apart from the original phase
space variables newly introduced variables were also used, we have
obtained the dynamical 2-form theory in the extended phase space.
We have also shown how reducibility of constraints appears in
this model. In Sec.~\ref{secf}, we have discussed the role played
by the auxiliary field $C_\m$ in the dynamical 2-form theory. Using
the phase space path integral approach, we have shown that the
dynamical 2-form theory cannot be obtained by a Hamiltonian
embedding of the na\"\i ve non-Abelian model.

It is of interest to note the difference in Poisson bracket
structures of the constraints of Freedman-Townsend model and the
dynamical 2-form theory. As we have seen the constraint algebra is
on-shell reducible in the case of former while in the case of
latter it is off-shell reducible.  In the case of Freedman-Townsend
model, we see that the Poisson brackets of the scalar constraints
$\L^a$ of Eqn.~(\ref{fof.Lafinal}) with all other constraints
vanish weakly.  On the other hand, the Poisson brackets of
$\L_{i}^a$ of Eqn.~(\ref{fof.Liafinal}), which are the generators
of the vector symmetry, with the remaining constraints vanish
strongly like in an Abelian theory. This shows that that the model
described by Eqn.~(\ref{intro.naive}) is not a pure non-Abelian
theory with respect to 2-form potential, unlike the Yang-Mills
gauge field.  We see the same feature in the case of dynamical
2-form theory also. Here the Poisson brackets of the generators of
SU(N) symmetry, as given in Eqn.~(\ref{sof.lamfin}), with other
constraints vanish weakly, while the Poisson brackets of the first
class constraint of Eqn.~(\ref{sof.Lambdafinal}) are strongly zero
like that of an Abelian theory. Thus we see here, in the
Hamiltonian formulation that the model described by
Eqn.~(\ref{intro.nalag}) is not a pure non-Abelian theory with
respect to the 2-form potential. This is expected as the vector
gauge symmetry of Eqn.~(\ref{intro.navsym}) is Abelian in the case
of both these models, as can be seen in the Lagrangian formulation.

It will be of interest to generalize the procedure of constraint
deformation applied here so that it can be applied to other models
also. Since both Freedman-Townsend model and dynamical 2-form 
theory also have second class constraints, it should be of interest to
elevate them also to first class ones either by BFT procedure or by
further deformation of constraints and study the corresponding gauge
theories. 

\noindent {\bf Acknowledgements}:

EH thanks the University Grants Commission, India for support.

\newpage

\appendix

\section{Gauge unfixing for the non-Abelian two-form}\label{a2}

Apart from the generalized canonical scheme developed by Batalin,
Fradkin and collaborators, there is another method to convert a
system with second class constraints to gauge theory. In this
method known as gauge
unfixing~\cite{{Mitra:1990mp},{Anishetty:1993yk}}, one modifies the
second class constraint and Hamiltonian using the original phase
space variables alone unlike that in the case of BFT embedding. In
this scheme, among the second class constraints
$T_\a,~(\a=1,...,2n)$, half of them $({\bar T}_a =T_\a ,\a=1,..n)$
are taken to be the constraints and the remaining half $(T_a,
\a=n+1,...2n)$ are taken as the corresponding gauge fixing
conditions.  Then using the constraints ${\bar T}_a$, a projection
operator is defined using which gauge invariant Hamiltonian and
other observables are constructed.

Here we apply the gauge unfixing to na\"\i ve non-Abelian model
described by Eqn..~(\ref{intro.naive}). The primary constraints
following from the Lagrangian of Eqn.~(\ref{intro.naive}) are
\begin{eqnarray}
\Pi_{0}^a\we0,~~~~\Pi_{0i}^a\we0,
\end{eqnarray}
where $\Pi_\m$ and $\Pi_{\m\n}$ are the momenta corresponding to
$A^\m$ and $B^{\m\n}$ respectively.  The total Hamiltonian is
\begin{eqnarray}
H_T&=& \fr{1}{4}\Pi_{ij}\Pi^{ij}
-\fr{1}{2}(\Pi_{i}^a-\fr{m}{2}\e_{0ijk}B^{ajk})
(\Pi^{ai}-\fr{m}{2}\e^{0ilm}B_{lm}^a) +\fr{1}{4}F_{ij}^a F^{aij}
-\fr{1}{12}H_{ijk}H^{ijk} \no\\
&-&A^{a0}\left [ (D^i\Pi_i)^a-\fr{1}{2}gf^{abc}\Pi_{ij}^b B^{cij}\right]
+B^{a0i} \left [ (D^j\Pi_{ij})^a-\fr{m}{2}\e_{0ijk}F^{ajk}\right],
\label{app.ham}
\end{eqnarray}
which can also be written as $H_T=H_c-A^{a0}\L^a -B^{a0i}\L_{i}^a,$
where the secondary constraints are 
\begin{eqnarray}
\L^a&=& (D^i\Pi_i)^a-\fr{1}{2}gf^{abc}\Pi_{ij}^b B^{cij},
\label{sec1}\\
\L_{i}^a &=& -(D^j\Pi_{ij})^a+\fr{m}{2}\e_{0ijk}F^{ajk}.
\label{sec2}
\end{eqnarray}
The constraint $\L^a$ does not lead to any further constraint as
$\PB{\L^a}{H_c}=0$ while the persistence of $\L_{i}^a$ gives a
tertiary constraint
\begin{eqnarray}
\PB{\L_{i}^a} {H_c}=\fr{1}{2}gf^{abc}F^{blm}H_{ilm}^c
+gf^{abc}\Pi^{bl}\Pi_{il}^c
+\fr{m}{2}gf^{abc}\e^{0lmn}\Pi_{il}^bB_{mn}^c\equiv \Psi_{i}^a.
\label{sec.ter}
\end{eqnarray}
The Poisson brackets of primary constraints with all the
constraints vanish and those among the remaining constraints are
\begin{eqnarray}
\PB{\L^a}{\L^b} &=& -gf^{abc}\L^c,~~\PB{\L^a}{\L_{i}^b} =
gf^{abc}\L_{i}^c,\no\\ \PB{\L^a}{\Psi_{i}^a} &=&
gf^{abc}\Psi_{i}^c,~~~ \PB{\L_{i}^a}{\L_{j}^b} = 0,\no\\
\PB{\L^{ai}}{\Psi_{j}^b} &=& -g^2f^{bcd}f^{aed}F_{jl}^e F^{cij}
-\fr{1}{2} g^2f^{bcd}f^{aed}F_{mn}^e F^{cmn}\d_{l}^i + g^2
f^{acd}f^{bce}\Pi^{dij} \Pi_{lj}^e \equiv {\cal A}_{l}^{abi}.
\label{cala}
\end{eqnarray}
{}From the constraint algebra we see that
$\Pi_{0}^a,~\Pi_{0i}^a~{\rm and}~\L^a$ are first class and
$\L_{i}^a~{\rm and}~\Psi_{i}^a$ are second class.

{}From the constraint structure it is clear that for a theory which
has invariance under the vector gauge transformation of
Eqn.~(\ref{intro.navsym}), $\L_{i}^a$ has to be a first class
constraint. Thus in applying the gauge unfixing procedure we take
$\L_{i}^a$ to be the first class constraint and $\Psi_{j}^b$ to be
the corresponding gauge fixing condition. The projection operator
used to construct the gauge invariant observables is then defined
as
\begin{eqnarray}
{\cal P}=\exp~-\int d^3 x \Psi^{ai}\L_{i}^a
\end{eqnarray}
A particular ordering is used such that when ${\cal P}$ acts on
functions of phase space variables, the gauge fixing condition
$\Psi_{i}^a$ should be outside the Poisson
bracket\cite{{Mitra:1990mp},{Anishetty:1993yk}}. 
Using this we construct the gauge unfixed Hamiltonian
\begin{eqnarray}
{\cal H}_{GU} &=& {\cal P}~ H_T\no\\ &=&H_T -\int d^3 y
~\Psi(y)^{ai} \PB{\L_{i}^a}{H_T} +\fr{1}{2}\int d^3y ~d^3z~
\Psi(y)^{ai}~ \Psi(z)^{bj}~\PB{\L_{i}^a}{\PB{\L_{j}^b}{H_T}}
- \cdots\,.
\end{eqnarray}
{}From the constraint algebra (\ref{cala}), we see that the higher
order terms in the above series vanish since $\PB{{\cal
A}_{l}^{abi}}{\Lambda^c_j} = 0$, and thus we get
\begin{eqnarray}
{\cal H}_{GU}= H_T -\Psi_{i}^a\Psi^{ai} -gf^{abc}\Psi_{i}^aA^{0b}\L^{ci}
+\fr{1}{2}\Psi_{i}^a\Psi^{bj} {\cal A}_{j}^{abi}
\label{guham}
\end{eqnarray}
But it is straight forward to check, using the constraint algebra
(\ref{cala}) that $\PB{\L_{i}^a}{{\cal H}_{GU}}$ is non-vanishing
and hence ${\cal H}_{GU}$ is not gauge invariant. Thus here we see
that the gauge unfixing method also fails to convert the
na\"\i ve non-Abelian model to a first class system.


\end{document}